# Seamless Infrastructure independent Multi Homed NEMO Handoff Using Effective and Timely IEEE 802.21 MIH triggers

Zohra Slimane, Mohamed Feham and Abdelhafid Abdelmalek

STIC Laboratory University of Tlemcen Algeria
*{ z_slimani, m_feham, a_abdelmalek }@mail.univ-tlemcen.dz*

## ABSTRACT

*Handoff performance of NEMO BS protocol with existent improvement proposals is still not sufficient for real time and QoS-sensitive applications and further optimizations are needed. When dealing with single homed NEMO, handoff latency and packet loss become irreducible all optimizations included, so that it is impossible to meet requirements of the above applications. Then, How to combine the different Fast handoff approaches remains an open research issue and needs more investigation. In this paper, we propose a new Infrastructure independent handoff approach combining multihoming and intelligent Make-Before-Break Handoff. Based on required Handoff time estimation, L2 and L3 handoffs are initiated using effective and timely MIH triggers, reducing so the anticipation time and increasing the probability of prediction. We extend MIH services to provide tunnel establishment and switching before link break. Thus, the handoff is performed in background with no latency and no packet loss while ping-pong scenario is almost avoided. In addition, our proposal saves cost and power consumption by optimizing the time of simultaneous use of multiple interfaces. We provide also NS2 simulation experiments identifying suitable parameter values used for estimation and validating the proposed model.*

## Keywords

*NEMO, multihoming, seamless handoff, IEEE 802.21, MIH triggers, path loss model, NS2*

## 1. INTRODUCTION

It is now possible to deploy, in moving networks such as vehicle and aircraft networks, applications implying communications with the infrastructure or with other moving networks while profiting surrounding heterogeneous wireless capacities of communication (e.g ieee 802.11, ieee 802.16, 3GPP, 3GPP2). The protocol NEMO Basic Support (BS) [1] was proposed by the IETF for supporting the mobility of moving networks. NEMO allows an entire IP network to perform a layer 3 (L3) handoff. Transparent service continuity is achieved using a mobile router for mobility management on behalf of the transported mobile network devices. Handoff performance plays a crucial role in QoS-sensitive applications and real-time services in heterogeneous networks. Although NEMO BS has the merit to allow as of today the deployment and the experimentation of no time constraints services without having to function in a degraded mode, its performance (high latency, high packet loss and high signaling cost) is thus clearly considered as suboptimal and is not appropriate for time constraints applications.

Therefore, there have already been a number of studies and a large set of optimizations that try to address these issues ([6]-[17]). The proposed solutions rely on the optimization of each component of the handoff, using cross layer design, network assistance, multihoming, etc. However, minimal reached values of handoff latency and packet loss still do not fill real time and QoS-sensitive applications requirements. Consequently, NEMO with the above optimizations is still not sufficient for such applications and further improvements or solutions are needed.

In this paper, we propose a new multihoming based NEMO handoff scheme achieving seamless connectivity (precisely with zero latency and zero packet loss). Our cross layer design uses timely and effective MIH triggers (such as Link_Going_Down, Link_switch_Imminent) and required handoff time based adaptative MIH command services such as Link_Configure_Thresholds. We provide proactive surrounding networks attachment, home registration, tunnel establishment and then tunnel switching if necessary just before Link Down event. With this manner of executing the anticipation, we increase the probability of prediction avoiding ping-pong scenario and we save also cost and power consumption.
The rest of the paper is organized as follows. Section 2 introduces the related works on NEMO optimizations. Section 3 provides NEMO handoff components analysis and numerical evaluation. Section 4 gives an overview of IEEE802.21 STD and MIH services. In Section 5, we describe the details of our proposal and associated algorithms. In Section 6, NS2 implementation and simulation results are presented, and the performance of the proposed scheme is discussed. Finally, conclusions are stated in Section 7.

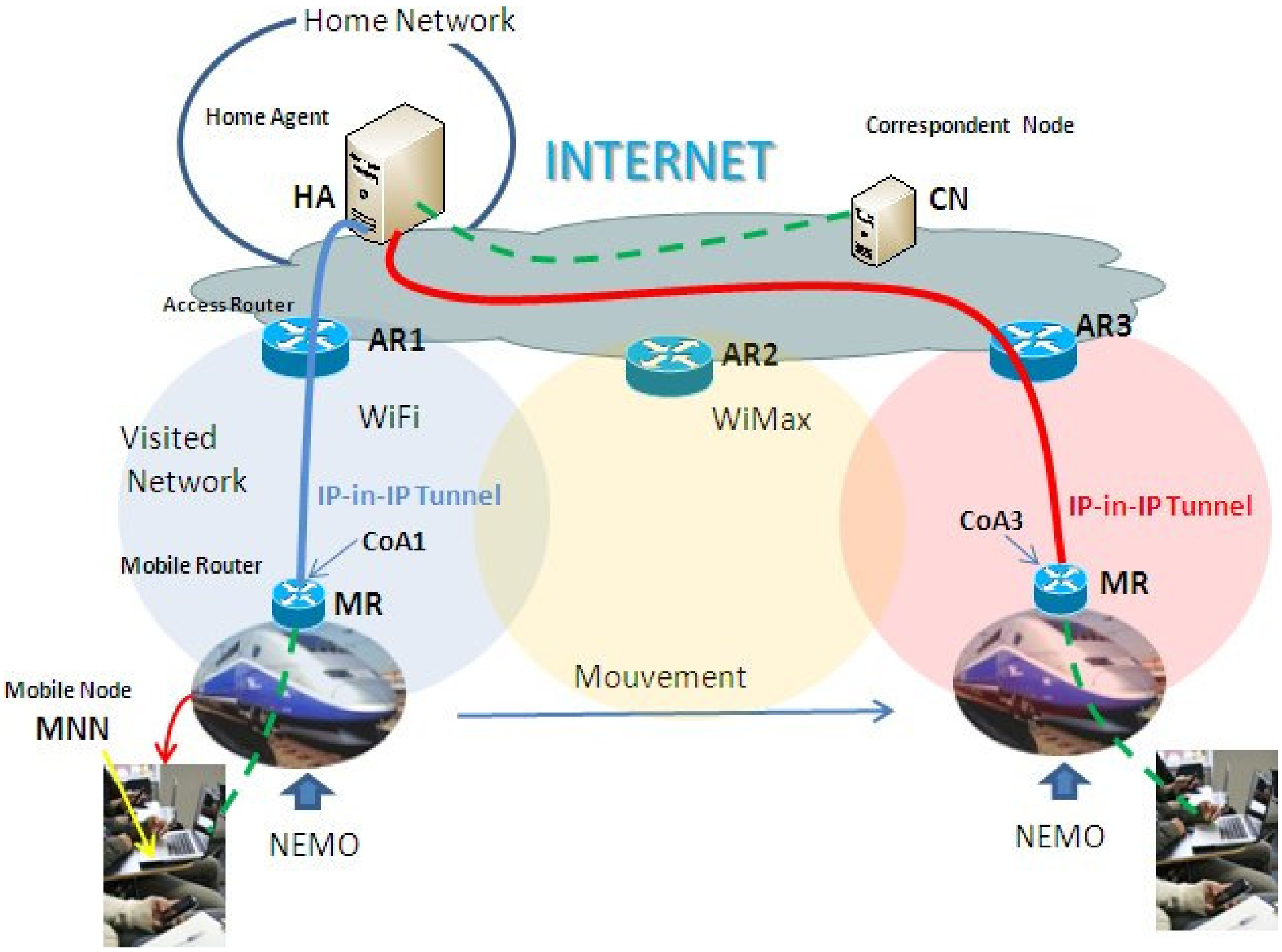

Figure 1. Basic Components of NEMO BS Protocol

# 2. Related Work

NEMO BS Protocol [1] designed by IETF to manage network mobility (Figure 1) is an extension of the MIPv6 [2] protocol. The NEMO BS (MIPv6-NEMO) handoff is composed of the link layer handoff followed by the new network attachment and then the home registration. Brake-Before-Make handoff performance (latency, packet loss and signaling overhead) of NEMO BS were analyzed in the literature [3, 4, 5, 8, 9, 18]. The results show that the mobility support does not provides seamless connectivity. To overcome the limitations of NEMO BS protocol, many optimizations were proposed. To reduce the new network attachment for MIPv6-NEMO, delay Optimistic Duplicate Address Detection (ODAD) [6] and Fast Router Advertisements [3,7] were proposed. Besides, Many Infrastructure based mobility supports were proposed to address handoff efficiency in NEMO. Cross layer design scheme [8] using IEEE 802.21MIH services and addressing movement prediction and handoff timing algorithms was proposed on FMIPv6 to anticipate L3 handoff. HiMIP-NEMO [9] proposes the use of Foreign Mobility Agent (FMA) to achieve QoS handoff with reduced latency and packet loss. An extension of Proxy MIPv6 (PMIPv6) called N-NEMO was proposed to provide mobility support for NEMO context [10, 21] . The scheme is based on tunnel splitting, global tunnel between LMA and MAG, and local tunnel between MR and MAG, leading to reduced signaling cost.

Many other works based on multihoming were investigated to improve seamless handoff. In [11] a new entity ICE (Intelligent Control Entity) is introduced in NEMO architecture to improve handoff for multiple MRs-based multihomed NEMO. Another protocol called SINEMO [12] using IP diversity and soft handoff was proposed to reduce Handoff signaling cost for a single multihomed MR based NEMO. Other GPS Aided Predictive Handover Management solutions using Make Before Brake handoff were proposed to improve handoff performance but they are rather more suitable for multihomed train-based NEMO [13, 14, 15 ].

Higher layer Extensions such as SIP-NEMO [16] and HIP-NEMO [17] based respectively on Session Initiation Protocol (SIP) and Host Identity Protocol (HIP) were also proposed. However, in addition to being not transparent to all applications these schemes suffer from additional signaling overhead.

The handoff performance of NEMO BS with above optimization is still not sufficient for QoS-sensitive applications. Latency of link layer handoff and NEMO signaling overhead (precisely from the round trip time RTT between the Mobile Router and the Home Agent) affect the overall performance of mobility management significantly.

# 3. NEMO handoff Latency Analysis

NEMO Basic Support (BS) protocol proposed by IETF provides mobility support for an entire mobile network moving across different heterogeneous access networks Continuous and uninterrupted internet access to the Mobile Network Nodes (MNN) inside the mobile network is provided by the Mobile Router (MR) which manages the movement (Figure 1). The MR is identified by its Home Address (HoA) through which it is accessible in its home network, and it is localized by its Care-of-Address (CoA) acquired at visited network. The Home Agent (HA) located at the home network assists the MR to support mobility management. To change its point of attachment to a new access network (i-e to a new access router AR), the MR must process in general a vertical Handoff including both L2 and L3 Handoff (Figure 2).

Since L2 and L3 Handoff are independent in NEMO BS protocol (L3 Handoff occurs after L2 Handoff), the overall handoff latency can be expressed by the following equation:

$$T_{HO} = T_{L2} + T_{L3} \tag{1}$$

Where $T_{L2}$ is the Link layer (L2) Handoff latency (the time required to establish a new association by the physical interface) and $T_{L3}$ is the IP layer (L3) Handoff latency (the time to register the new CoA at the Home Agent (HA) and to be able to receive the first data packet at this new localization).

L2 Handoff procedure includes in general scanning ($T_{scan}$), authentication ($T_{auth}$) and association ($T_{ass}$) which are very dependent on technology and exhibit great variation. The published values of $T_{L2}$ are between 50 ms and 400 ms [4, 19, 20].

Then:

$$T_{L2} = T_{scan} + T_{auth} + T_{ass} \tag{2}$$

The L2 Handoff is triggered by the link event:

$$P_{rx} < P_{th} \tag{3}$$

Where $P_{rx}$ is the received signal power corresponding to the received signal strength indication (RSSI) and $P_{th}$ is the predefined threshold power below which the Link status is considered down.

L3 Handoff procedure is composed of four distinct phases:

- Movement Detection (MD): after disconnecting from the old AR (oAR), the MR detects its movement thanks to prefix information contained in received Router Advertisement (RA) messages broadcasted periodically by the new AR (nAR). The MR may proactively send Router Solicitation (RS) messages to obtain the RA message from the nAR (The MR detects its movement if the oAR is unreachable, i-e no RA messages from the oAR).
- Duplicate Address Detection (DAD): Upon receiving prefix information from the nAR, the MR proceeds to the stateless auto-configuration; it configures itself with a new CoA (constructed from new prefix) and must check its uniqueness with the DAD process.

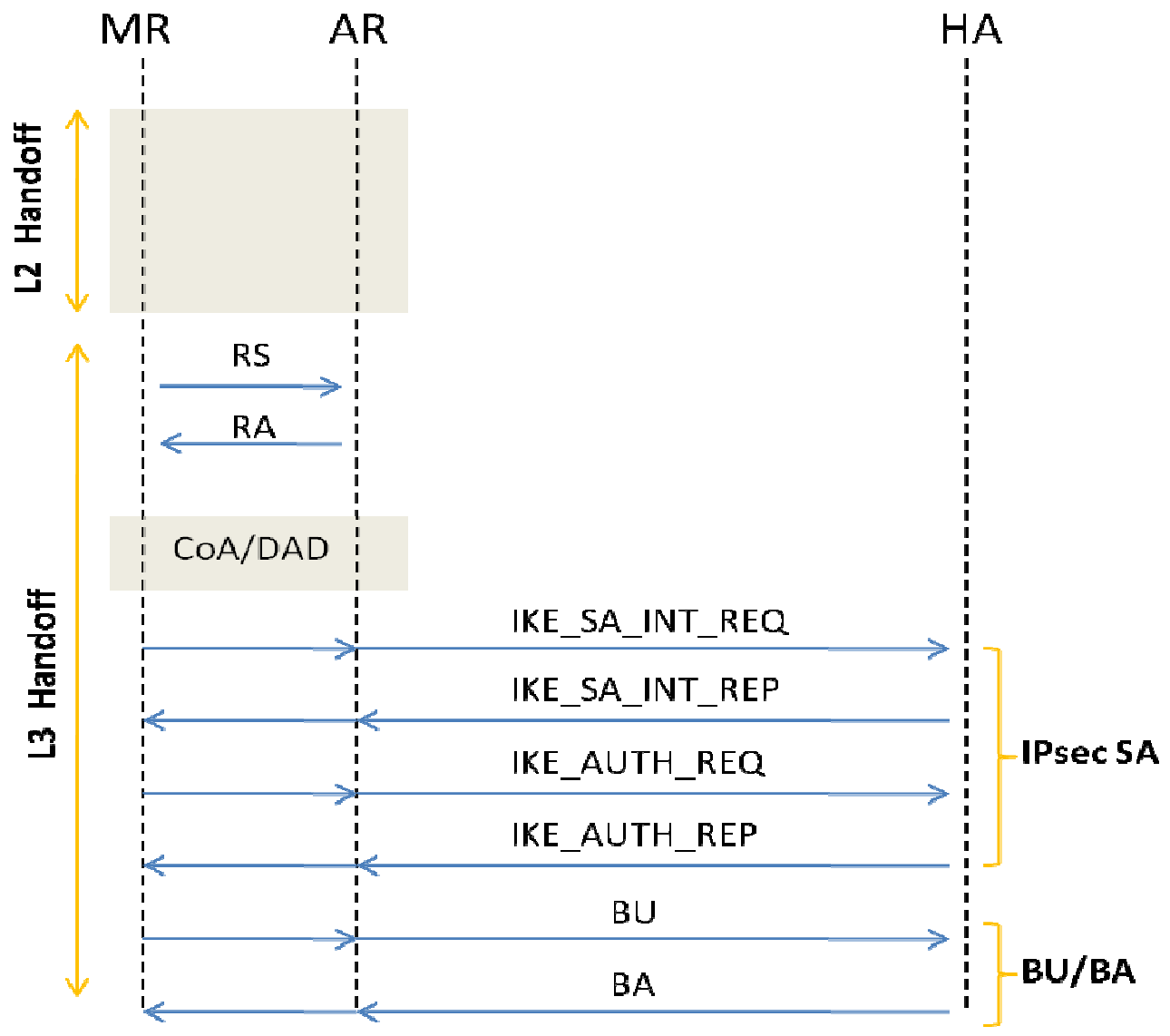

Figure 2. NEMO BS Protocol handoff procedure

- New CoA Registration and MR-HA Tunnel establishement (Reg): As soon as the MR acquires a new CoA, it immediately sends a Binding Update (BU) to its Home Agent (HA). Upon receiving this message, the HA registers the new CoA in its binding cache and acknowledges by sending a Binding Acknowledgement (BA) to the MR. As stated by [1], all signaling messages between the MR and the HA must be authenticated by IPsec. Once the binding process finishes, a bi-directional IP-in-IP tunnel is established between the MR and its HA. The tunnel end points are the MR's CoA and the HA's address. Either IPsec or other IP-in-IP protocol could be used for this purpose.

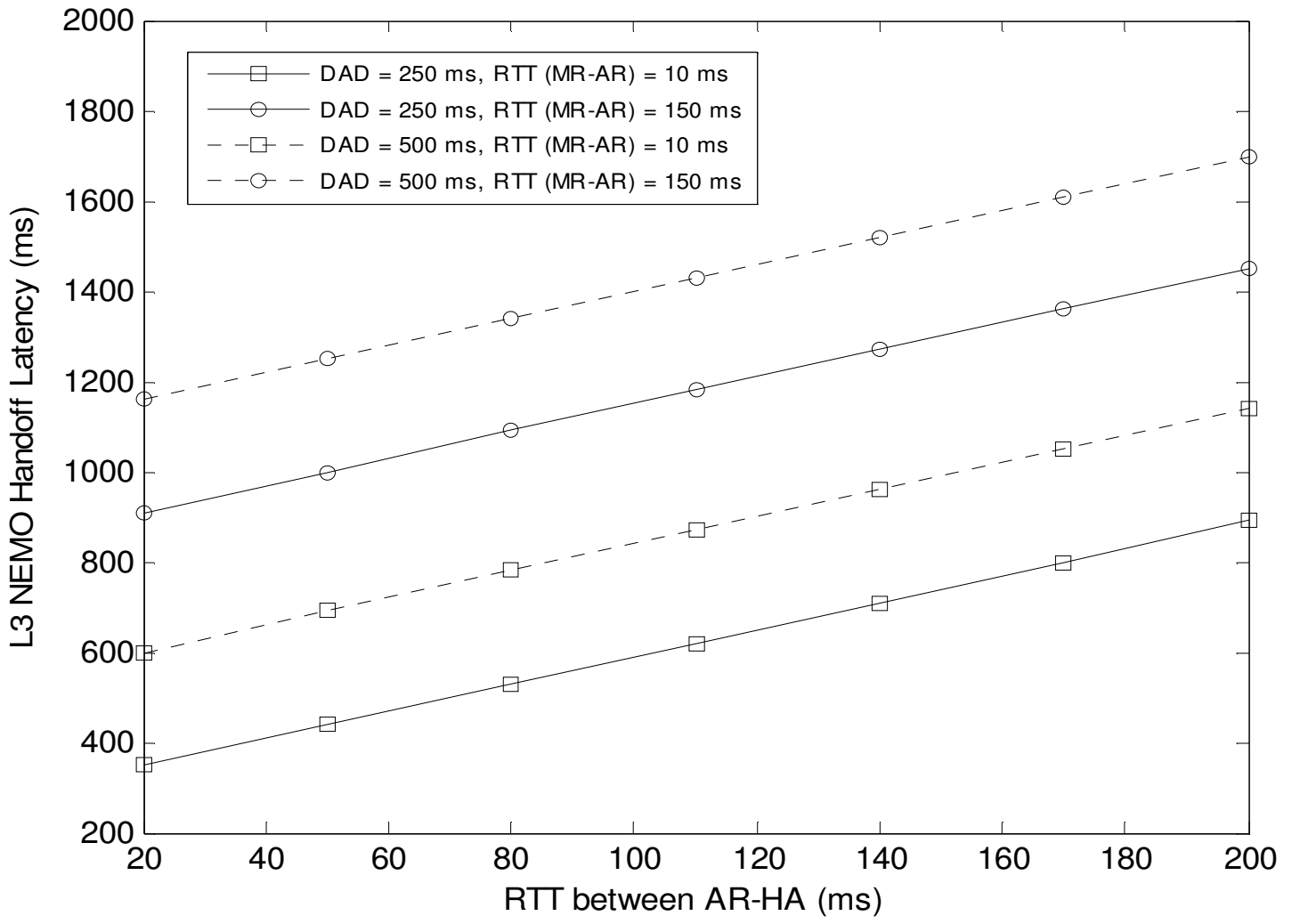

Figure 3. L3 NEMO Handoff latency vs. $RTT_{AR-HA}$

Thus, The L3 Handoff latency can analytically be computed as:

$$T_{L3} = T_{MD} + T_{DAD} + T_{Reg} \quad (4)$$

Where $T_{MD}$ , $T_{DAD}$ and $T_{Reg}$ are respectively Movement Detection phase delay, DAD process delay and registration delay.
Additionally, we have in the explicit form:

$$T_{MD} = T_{RS} + T_{RA} \quad (5)$$

$$T_{Reg} = T_{SA} + T_{BU} + T_{BA} \quad (6)$$

Where:
$T_{RS}$ : delay of Router Solicitation
$T_{RA}$ : delay of Router Advertisement
$T_{SA}$ : delay of creating an IPsec Security Association (SA)
$T_{BU}$ : delay of Binding Update
$T_{BA}$ : delay of Binding Ack

Then, according to (Figure 2) we can compute $T_{L3}$ as function of $RTT_{MR-AR}$ and $RTT_{AR-HA}$ , where RTT is the Round Trip Time.

$$T_{L3} = 4\, RTT_{MR-AR} + T_{DAD} + 3RTT_{AR-HA} \quad (7)$$

(Figure 3) and (Figure 4) show respectively L3 NEMO Handoff Latency and Overall NEMO Handoff Latency (L2+L3). For $RTT_{MR-AR}$ , we use a minimum value of 10 ms and a maximum value of 150 ms. For $RTT_{AR-HA}$ (twice time the delay of internet) we use the measured data from [22].

Two values of DAD (250, 500 ms) are used to take account of optimistic DAD. We can easily see that the minimum value of the Total NEMO Handoff Latency exceeds 400 ms, and this minimum values are carried out only under very special conditions.

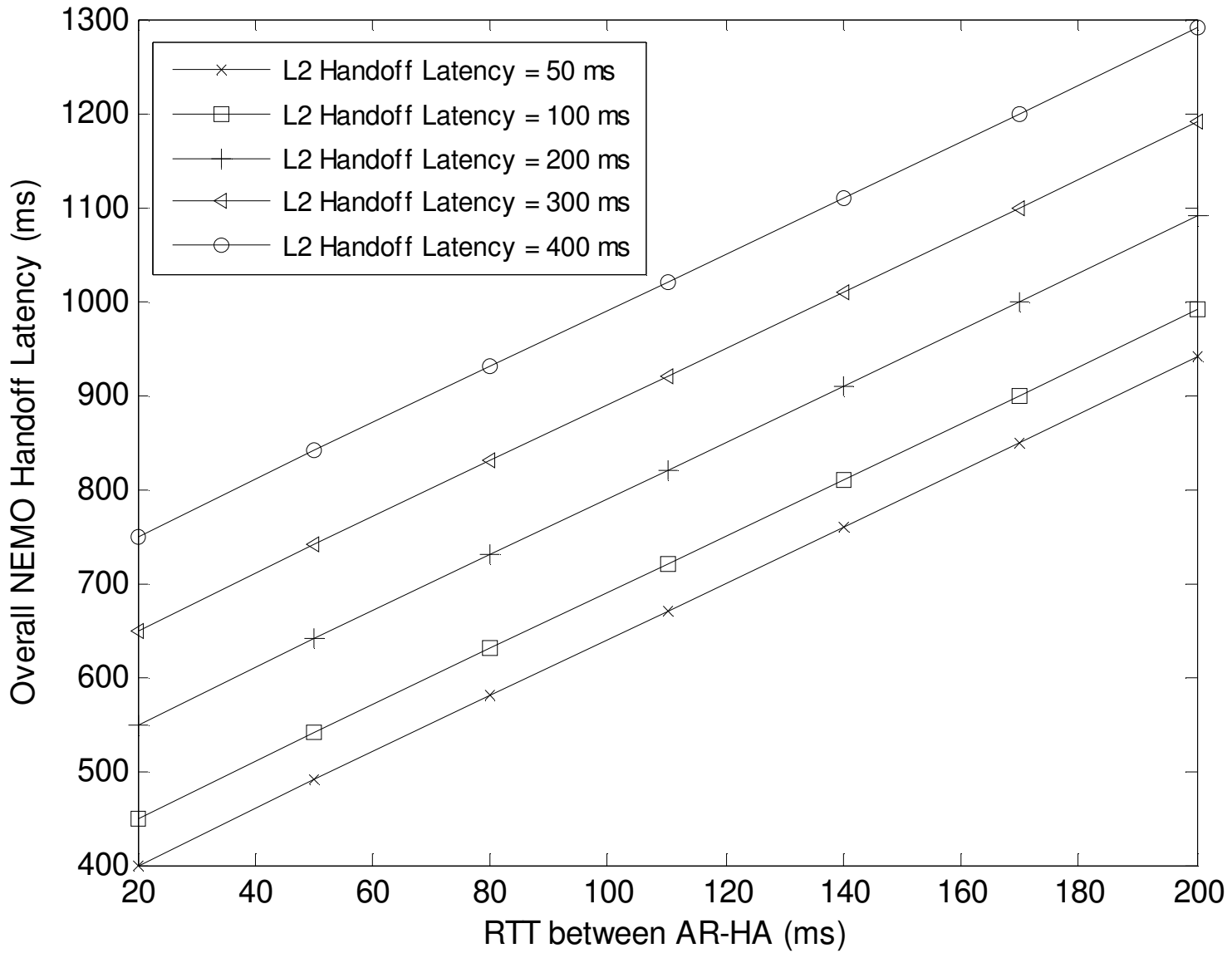

Figure 4. Overall NEMO Handoff latency vs. $RTT_{AR-HA}$
(only the minimum value 10 ms of $\mathrm{RTT_{MR-AR}}$ is considered)

(Figure 5) shows the Packet Loss during Handoff increasing with both the overall NEMO Handoff latency and the data rate. The results provided by [8] for example for NEMO Handoff improvements experienced for vehicular networks based on MIH assisted FMIPv6 show an overall NEMO Handoff latency of about 250 ms when vehicle has a slow movement (18 Km/s) and this value increases to 350 ms when vehicle speed reaches 90 Km/h. Consequently, these results show that single homed NEMO even improved is not appropriate for real time and QoS-sensitive applications.

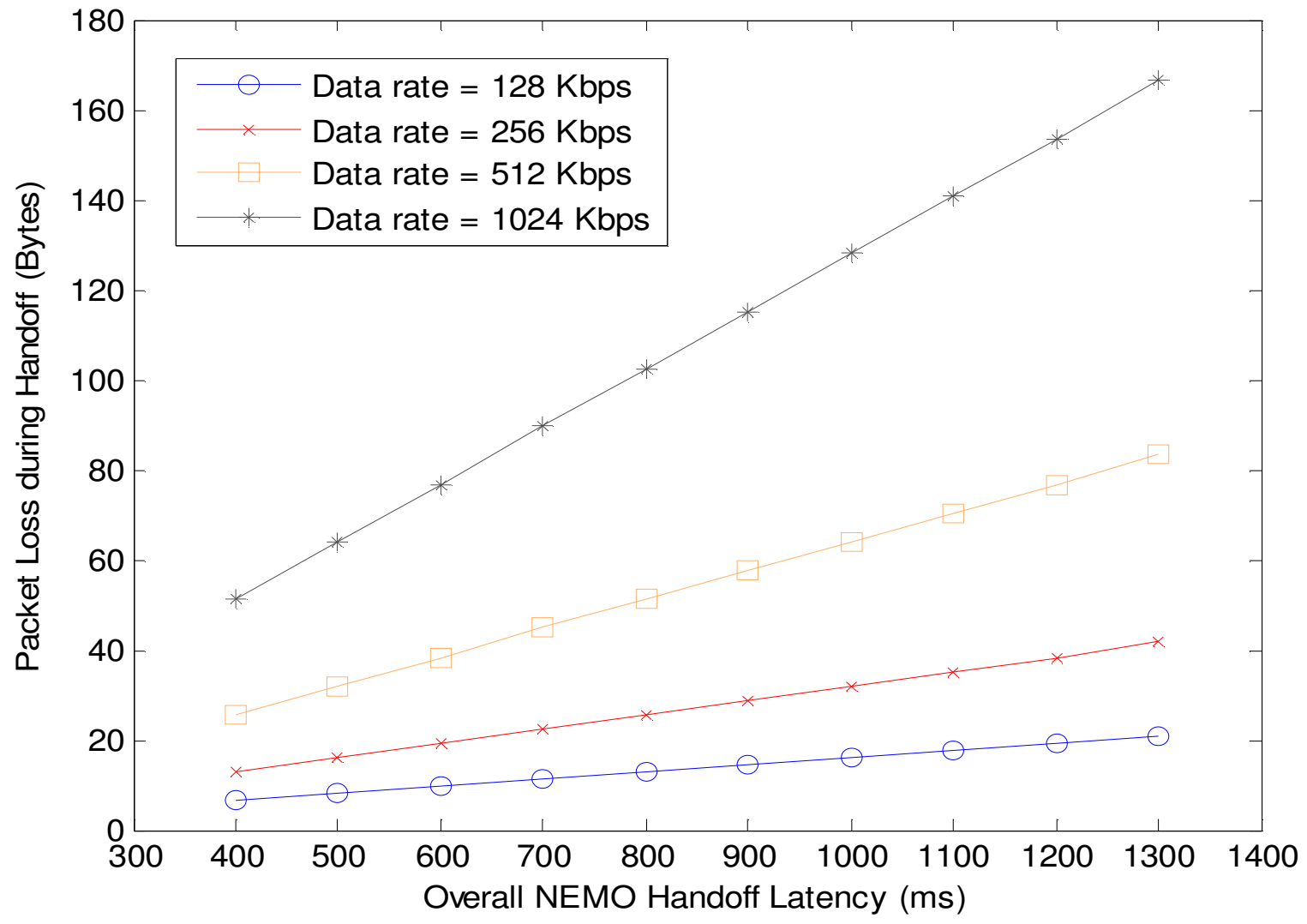

Figure 5. Packet loss during NEMO Handoff

# 4. IEEE 802.21 Media Independent Handover Services

The main aim of the IEEE 802.21 MIH standard [23] is the specification of generic SAPs and primitives that provide generic link layer intelligence and some network information to upper layers to optimize handovers between heterogeneous media such as IEEE 802.11 a/b/g/n, IEEE 802.16, 3GPP/3GPP2 etc. IEEE 802.21 provides a framework (a logical interface) that allows higher levels (users in the mobility-management protocol stack) to interact with lower layers to provide session continuity without dealing with the specifics of each technology.

## 4.1. MIH architecture

The core element of the MIH architecture is the MIH Function (MIHF) which is a logical interface between L2 and higher layers (Figure 6). MIHF which can be seen as a L2.5 layer helps in handover decision making and link selection by L3 and Upper layers by providing them with abstracted services. Upper layers (including mobility manager such as MIPv6 and NEMO, IP, transport protocols and applications) are the MIH Users. The MIH Users communicate with the MIHF via MIH_SAP (a media independent Service Access Points). The MIHF, on the other hand, interacts with L2/L1 layers via the MIH_LINK_SAP.

## 4.2. MIHF services

MIHF defines three main services that facilitate handovers between heterogeneous networks: MIH Event Services (MIES), MIH Command Services (MICS) and MIH Information Services (MIIS).

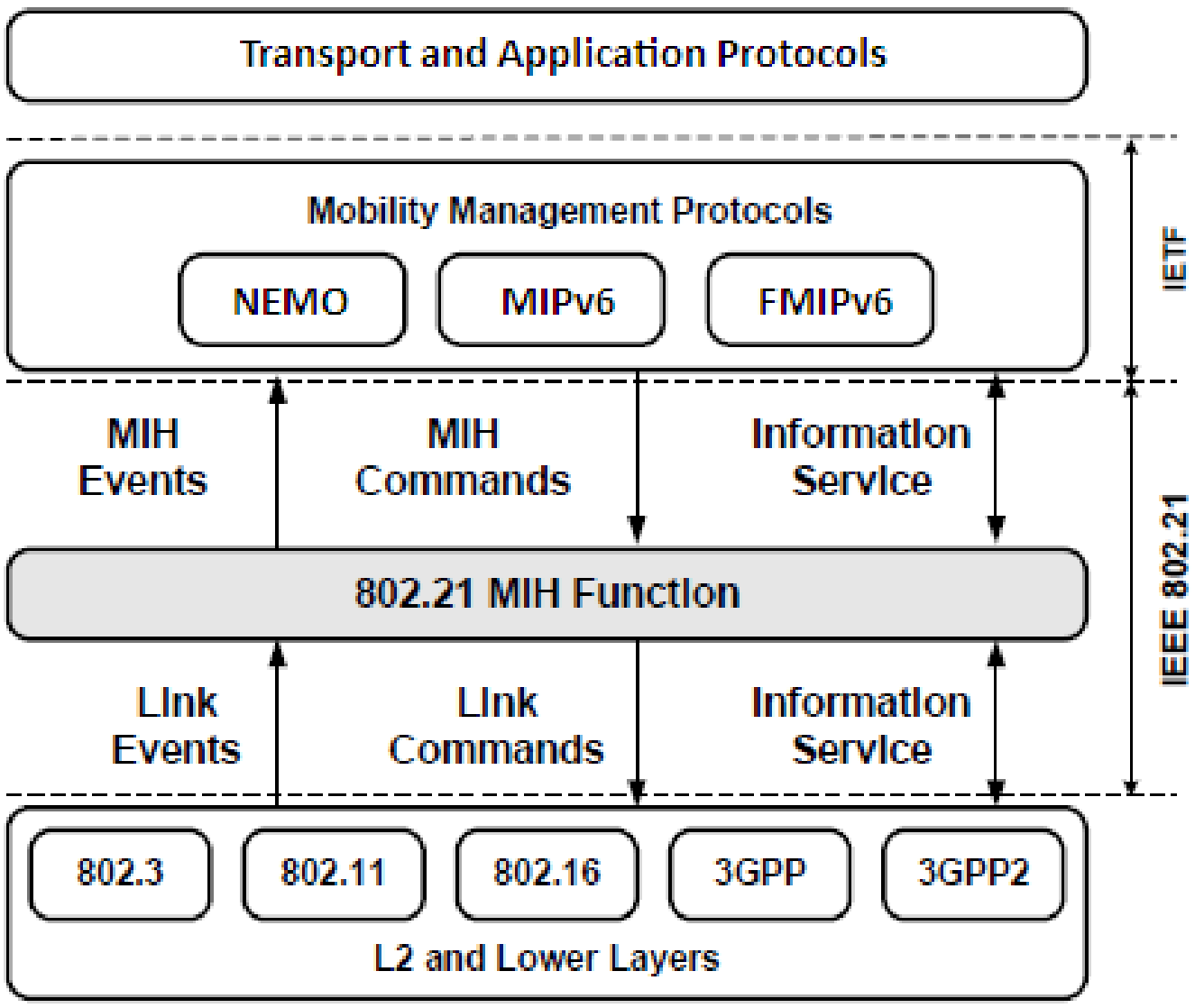

Figure 6. IEEE 802.21 General Architecture

#### 4.2.1. MIES - Media Independent Event Service

MIH capable devices use MIES to generate L1/L2 events indicating state and parameters changes occurring on the link to the upper layers. Two types of events are possible: Link Events exchanged between L1/L2 layers and the MIHF, and the MIH Events between the MIHF and the MIH Users. The defined events include Link Detected, Link Up, Link Down, Link Going Down, Link Parameters Change, Link Event Rollback, etc.

#### 4.2.2. MICS - Media Independent Command Service

MICS are commands ordered by the MIH Users to the lower layers to control their behavior. Two types of commands are possible: Link Commands issued by the MIHF to the lower layers such as Link Configure Thresholds and MIH Commands issued by MIH Users to the MIHF such as Get status, Switch, Configure, Configure Link Thresholds, Scan, Handover Initiate, Handover Terminate, etc.

#### 4.2.3. MIIS - Media Independent Information Service

MIH Users rely on The MIIS to obtain information from remote MIHF about available access networks. Potential target networks and their capabilities could be discovered to facilitate handovers by making more accurate decisions. MIIS includes support for various Information Elements (IEs) which includes information about network such as Identifier, cost, QoS and security, and information about Point of Attachment (PoA) such as location, Link-layer address, subnet, data rate, etc.

## 5. Proposed MIH assisted Multihomed NEMO Handoff

In this section, we will describe our proposed scheme for managing mobility with NEMO when a multihomed MR is used. First, we present our model, then we explain the MIH services to be used, and finally the procedure of Handoff is detailed.

## 5.1. Mobility Management Model

In our proposed, we suppose a (1,1,1) Multihomed NEMO model (i-e: one MR, one HA and one MNP-Mobile Network Prefix) [27].We consider so a mobile network with a single MR integrating multiple interfaces. These interfaces should be from different technologies or from same technology. Duplicate interface will be used in soft handoff to gain access to new network using same technology as current network becoming unreachable. The MR has a unique HoA and may obtain different CoA simultaneously. The MR is IEEE 802.21 compliant, and to provide an infrastructure independent scheme only local MIH services are used. Therefore, the HA must support multiple CoA (MCoA) registration [28]. Our scheme relies on three entities in the mobility management stack: MIHF, the Handoff Policy Decision entity (HPD) and NEMO protocol, the two last entities are MIH Users.

## 5.2. MIH services used in our Scheme

We utilize a subset of existing MIH services and new proposed ones to facilitate handoff decision making. (Table 1) lists these services (primitives) with corresponding parameters.

Table 1. Used MIH services in the proposed approach.

| Primitive | Service | Parameters |
|---|---|---|
| MIH_Link_Detected | MIES | MR IF MAC Addr, MAC addr of new PoA, MIH capability, Link Type |
| MIH_Link_Up | MIES | MR IF MAC Addr, MAC addr of new PoA, Link ID |
| MIH_Link_Down | MIES | MR IF MAC Addr, MAC addr of new PoA, Reason Code |
| MIH_Link_Going_Down | MIES | MR IF MAC Addr, MAC Addr of Curent PoA, TimeInterval, ConfidenceLevel |
| MIH_Link_Switch_Imminent | MIES (new) | MR IF MAC Addr, MAC Addr of Curent PoA, TimeInterval, ConfidenceLevel |
| MIH_Link_Event_Rollback | MIES | MR IF MAC Addr, Event ID |
| MIH_Configure_Link_Threshold | MICS | LinkParameter, nitiateActionThreshold, RollbackActionThreshold, ExecuteActionThreshold |
| MIH_Switch | MICS | Old Link ID, New Link ID |

When using a single interface, the MR cannot be associated simultaneously with more than one AR. Therefore, it has to break its communication with its current AR (hard handoff) before establishing an association to a new one. Hence, the handoff process is triggered by the Link_Down (LD) event. In our proposed scheme based on multihoming, Handoff process should be finished before the Link_Down event of the current link. So, instead of using LD trigger, we provide Link_Going_Down (LGD) and Link_Switch_Imminent (LSI) events which are fired using required Handoff time and required tunnel switching time. (Figure 7) shows corresponding received power threshold (RSS) of each event.

$\alpha_{LGD}$ and $\alpha_{LSI}$ are respectively the LGD power level threshold coefficient and the LSI power level threshold coefficient ($\alpha_{LGD} > \alpha_{LSI} > 1$). We use LGD event to trigger a soft handoff, and LSI event to switch tunnel before LD event. LSI event is used also to increase the probability of prediction and to avoid ping-pong scenario.

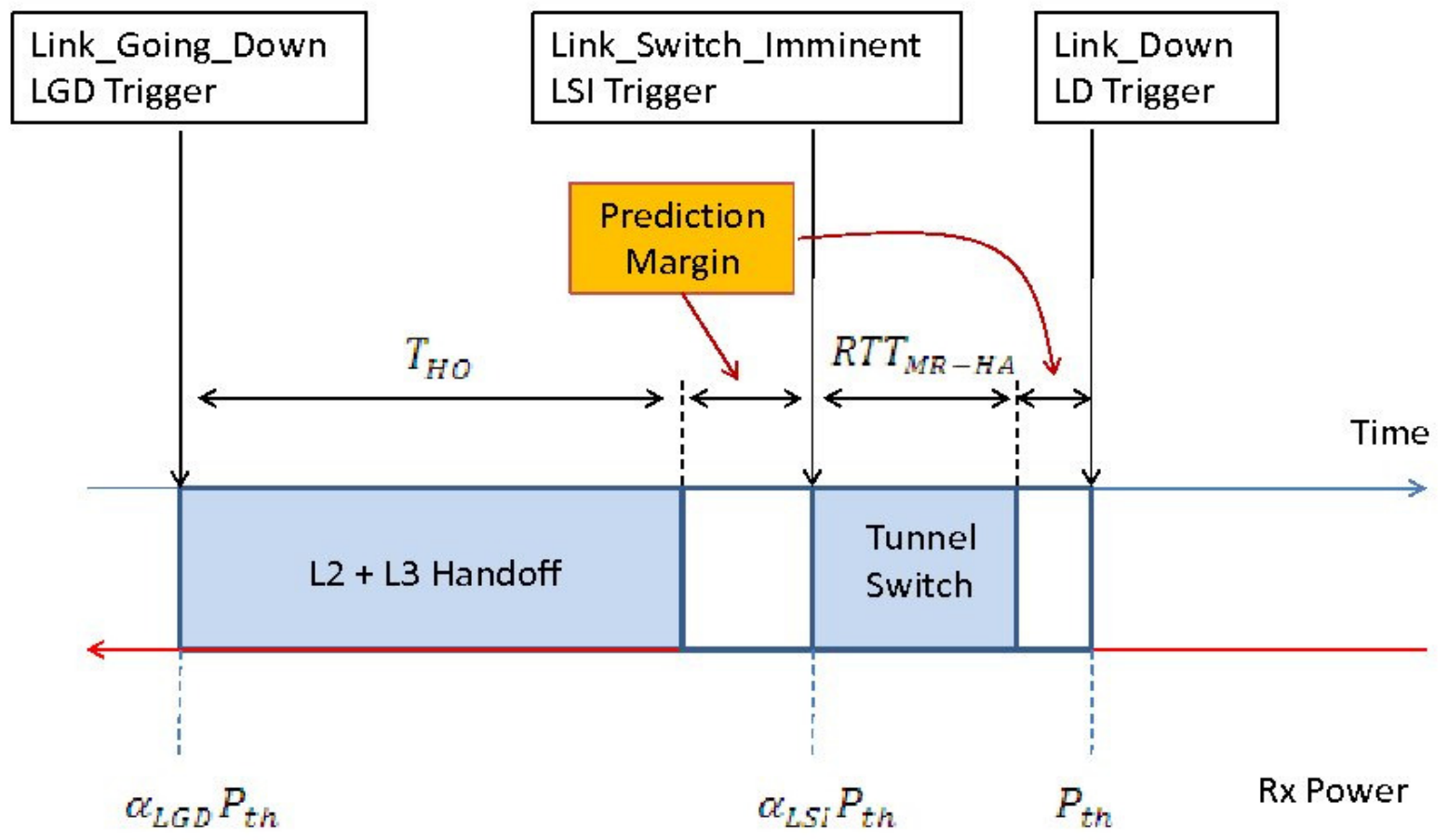

Figure 7. Generated Link triggers to prepare and perform Handoff before Link_Down event

LGD trigger in [23] is based on pre-defined threshold associated with the received signal strength (RSS). If the measured value of RSS crosses threshold $\alpha_{LGD}P_{th}$ , then the LGD trigger is generated and the handover process starts.

In our proposal $\alpha_{LGD}$ and $\alpha_{LSI}$ coefficients are adaptively configured using information gathered from neighboring access networks (we use for this purpose MIH_Configure_Link_Threshold primitive).

## 5.3. Required Handoff Time and Tunnel switching Time Estimation

The required handoff time $T_{HO}$ and tunnel switching time $T_{TS}$ are important factors for timely link triggering. The LGD trigger should be invoked prior to an actual LD event by at least the time required to prepare and execute a handoff. LSI trigger should be generated $T_{TS}$ before LD event. In our scheme, the setting $\alpha_{LGD}$is based on the following total time $T_{LGD}$:

$$T_{LGD} = T_{HO} + \Delta T_{HO} + T_{TS} + \Delta T_{TS} \quad (8)$$

Where :

$T_{HO}$ is given by (1)

$\Delta T_{HO}$ and $\Delta T_{TS}$ are added as security margin.

$$\Delta T_{HO} = \gamma_1 \% T_{HO} \quad (9)$$

$$\Delta T_{TS} = \gamma_2 \% T_{TS} \quad (10)$$

$\gamma_1$ and $\gamma_2$ are between 0 and 20.

Equation (10) can be written in the following form:

$$T_{LGD} = T_{L2} + T_{L3} + \Delta T_{HO} + RTT_{MR-HA} + \Delta RTT_{MR-HA} \quad (11)$$

In the same way, we get for $\alpha_{LSI}$ :

$$T_{LSI} = RTT_{MR-HA} + \Delta RTT_{MR-HA} \quad (12)$$

To estimate (L2+L3) handoff time and tunnel switching time, we use:

- New detected link to get L2 handoff time estimation and $RTT_{MR-nAR}$ based on link type information.

- Current link to get L3 handoff time estimation and tunnel switching time estimation by measuring $RTT_{MR-oAR}$.

## 5.4. Setting LGD and LSI triggers Thresholds

Given a path loss model, an analytical method can be used for effectively setting $\alpha_{LGD}$ and $\alpha_{LSI}$ coefficients [24, 25]. Let's assume the log-distance path loss model [26] for example shown in (13).

$$\left[\frac{P_{rx}(d)}{P_{rx}(d_0)}\right]_{dB} = -10\beta\log\left(\frac{d}{d_0}\right) \quad (13)$$

where $d$ is the distance between the receiver and the transmitter expressed in meters, $P_{rx}(d)$ denotes the received signal power level in watts at distance $d$ , $\beta$ is the path loss exponent, and $P_{rx}(d_0)$ is the received power at the close-in reference distance , $d_0$, and can be determined using the free space path loss model (take for example $d_0 = 1\ m$).

Assuming the Mobile Network (NEMO) moving at speed $v$, then $\alpha_{LGD}$ and $\alpha_{LSI}$ coefficients can be determined as:

$$\alpha_{LGD} = \left[\frac{1}{1-\frac{vT_{LGD}}{d_0}\left(\frac{P_{th}}{P_{rx}(d_0)}\right)^{\frac{1}{\beta}}}\right]^{\beta} \quad (14)$$

$$\alpha_{LSI} = \left[\frac{1}{1-\frac{vT_{LSI}}{d_0}\left(\frac{P_{th}}{P_{rx}(d_0)}\right)^{\frac{1}{\beta}}}\right]^{\beta} \quad (15)$$

Figures 8 and 9 respectively 10 and 11 show $\alpha_{LGD}$ and $\alpha_{LSI}$ variations for different $\beta$ values and different moving speeds. Both $\alpha_{LGD}$ and $\alpha_{LSI}$ increase with $\beta$, $v$ and required time for their setting. For example, we plot in Figure 12 the $\alpha_{LGD}$ variations versus $\beta$ for a mean value of $T_{LGD}$ equal to 1.25 s.

Note that speed $v$ can be estimated using the following approach:
Assume that at instant time $t_i$ the received signal power level is $P_{rx}(d_i)$ and at $t_{i+1}$ we receive $P_{rx}(d_{i+1})$, from (13) we get:

$$v = \frac{d_{i+1} - d_i}{t_{i+1} - t_i} \tag{16}$$

Therefore:

$$v = \frac{d_0}{t_{i+1} - t_i}\left[\left(\frac{P_{rx}(d_0)}{P_{rx}(d_{i+1})}\right)^{\frac{1}{\beta}} - \left(\frac{P_{rx}(d_0)}{P_{rx}(d_i)}\right)^{\frac{1}{\beta}}\right] \tag{17}$$

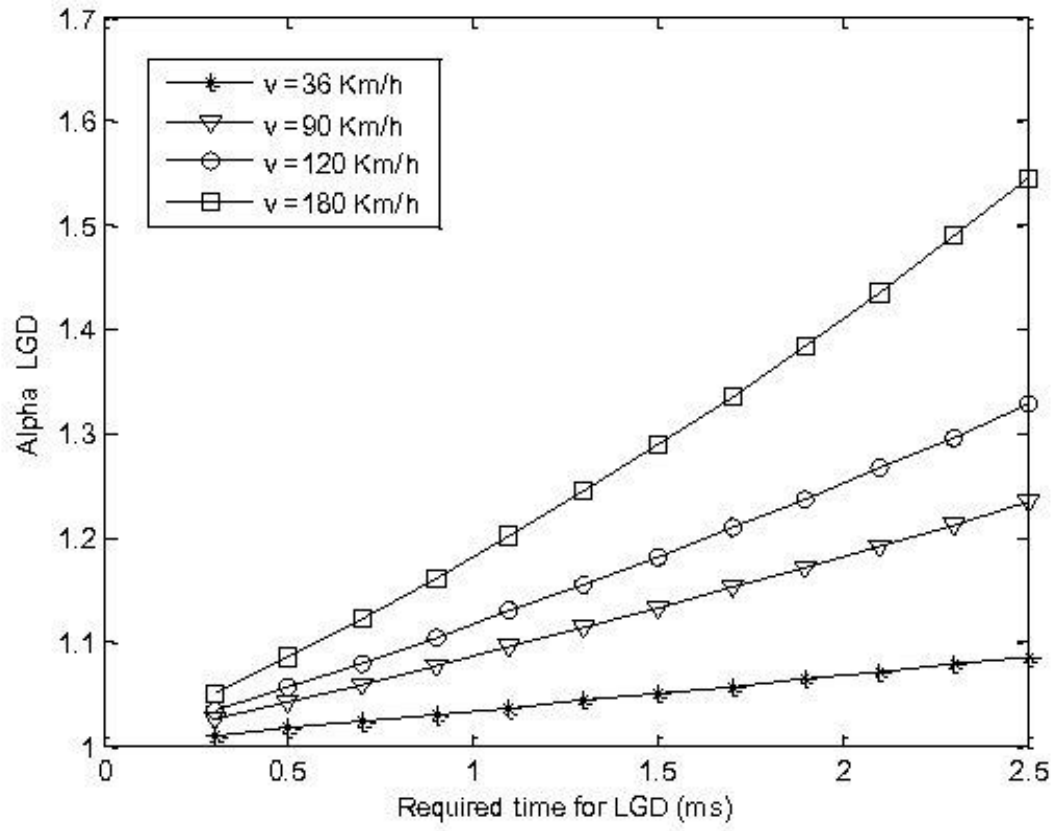

Figure 8. $\alpha_{LGD}$ vs. $T_{LGD}$ (for $\beta = 3$)

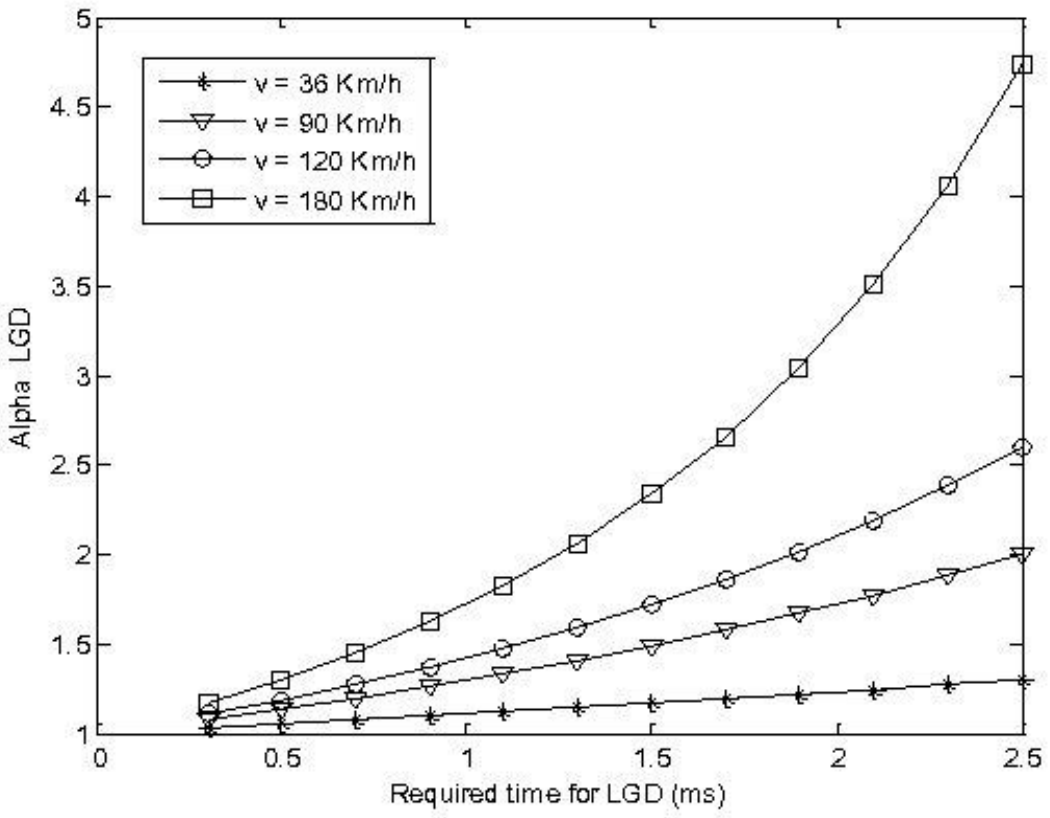

Figure 9. $\alpha_{LGD}$ vs. $T_{LGD}$ (for $\beta = 3.5$)

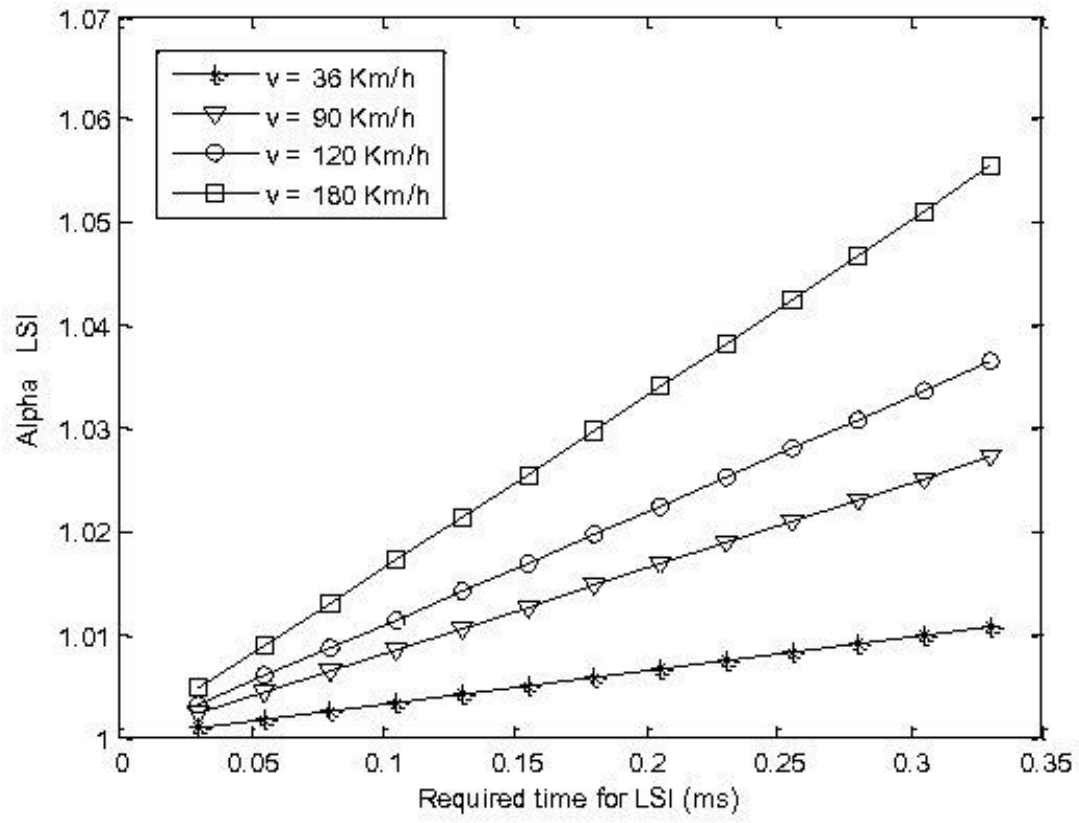

Figure 10. $\alpha_{LSI}$ vs. $T_{LSI}$ (for $\beta = 3$)

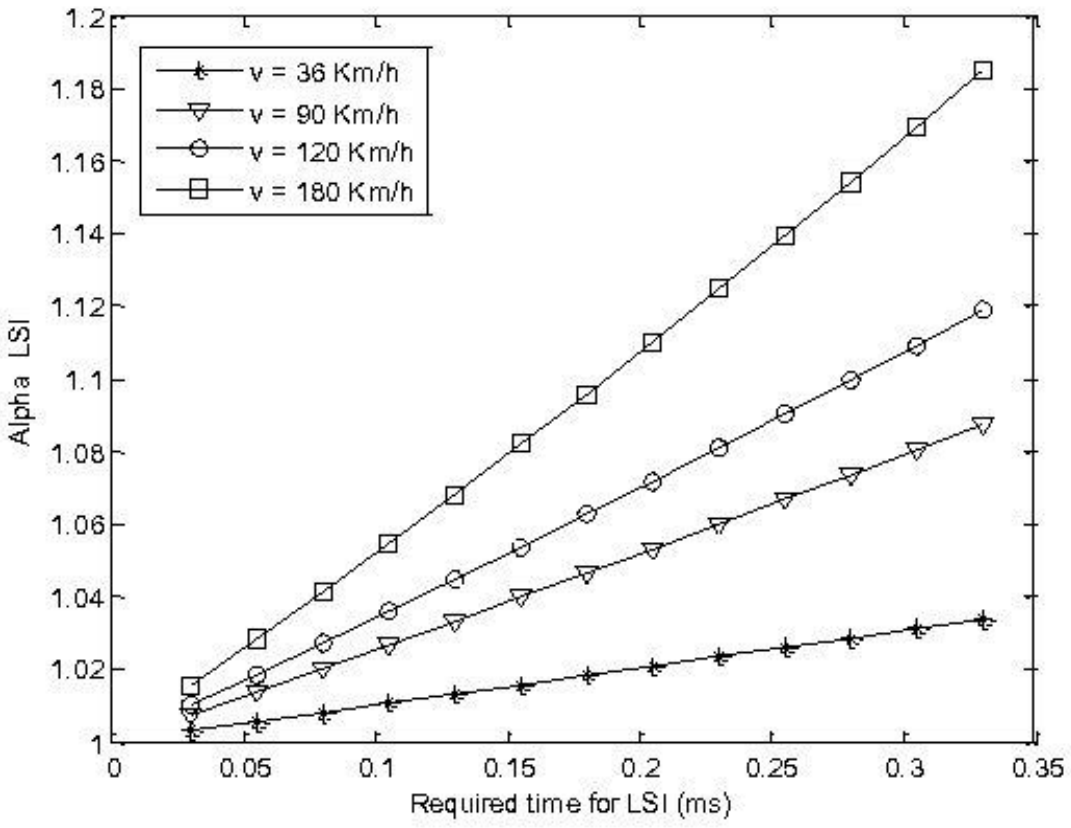

Figure 11. $\alpha_{LSI}$ vs. $T_{LSI}$ (for $\beta = 3.5$)

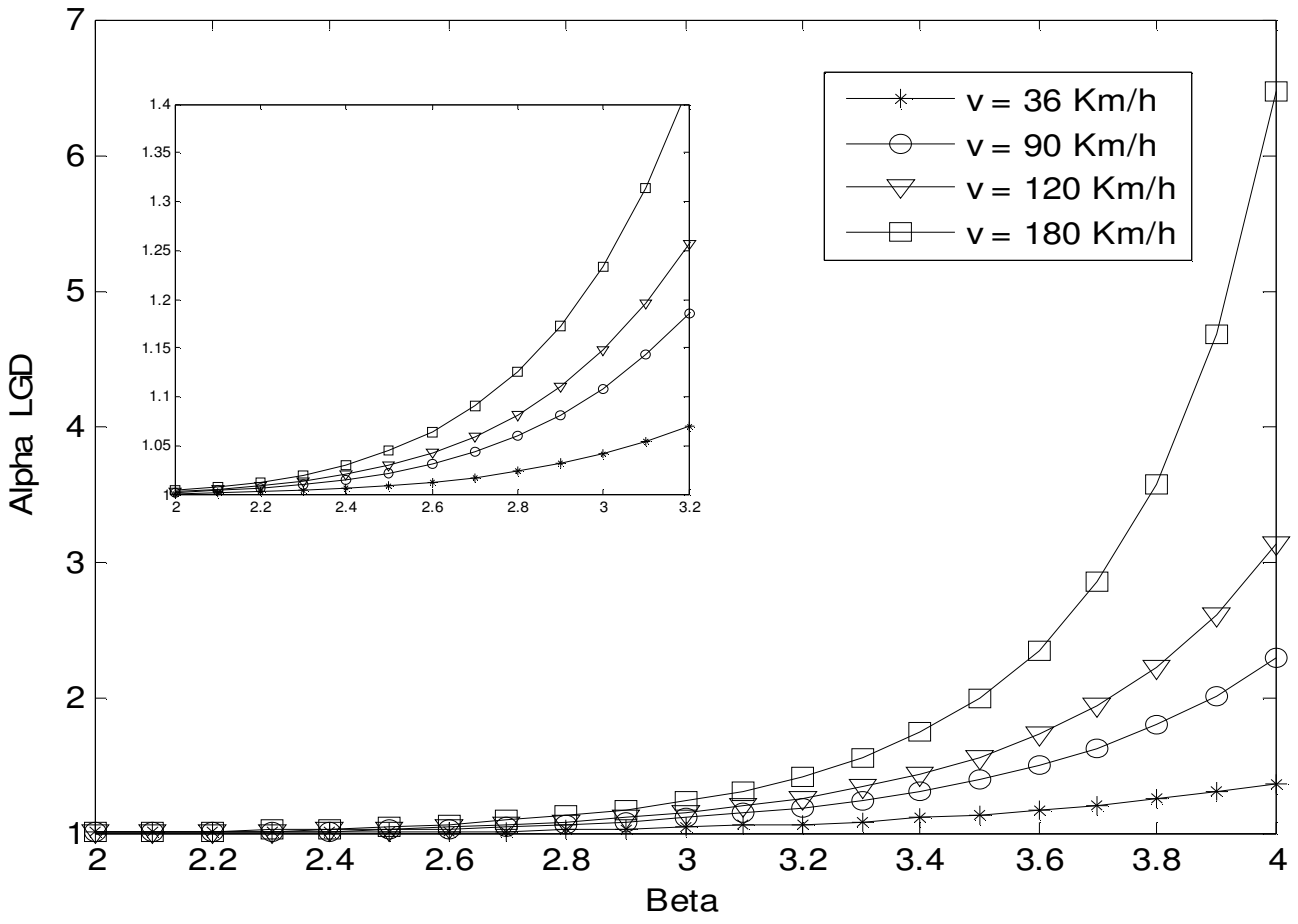

Figure 12. $\alpha_{LGD}$ vs. $\beta$ (for $T_{LGD} = 1.25\ s$)

However, to achieve a more realistic path loss model we have to take into account the shadowing effects which may affect the propagation model. An additional component $X_\sigma\ (dB)$ is introduced in the log-distance path loss model shown in (13) leading to the model known as the log-normal shadowing [26]:

$$\left[\frac{P_{rx}(d)}{P_{rx}(d_0)}\right]_{dB} = -10\beta \log\left(\frac{d}{d_0}\right) + X_\sigma \tag{18}$$

$X_\sigma$ is a zero-mean Gaussian distributed random variable with a standard deviation of $\sigma$.

When the shadowing component becomes significant, it is important to include a weighted averaging mechanism to produce a stable signal strength measure. We use for this purpose a simple recursive estimator:

$$\overline{P_{rx}}\ (i) = \delta\ P_{rx}(i) + (1-\delta)\overline{P_{rx}}\ (\text{i-1}) \tag{19}$$

where $\overline{P_{rx}}\ (i)$ is the average received signal power at instant i, $P_{rx}(i)$ is the received signal power at instant i and $\delta$ is the weighting factor.

## 5.5. Handoff operation and Tunnel switching

We suppose that the mobile network (NEMO) is already connected to an access network, and that a tunnel is already operational between the HA and the MR through one of its multiple interfaces. Let's denote this active interface IF-1. When the MR moves it could be covered by another access network. So, if a Link_Detected event is generated, by another interface (say IF-2), the MIHF translate this event to the HPD (Figure 13). This latter maintains a cache for detected links called *AvailableLinkCache* (Table 2).

So, when the HPD receives the MIH_Detected_Link event, it updates its cache and requests MIHF to generate MIH_Configure_Link_Threshold to set LGD and LSI triggers Thresholds for IF-1. Then, if a Link_Going_Down event is generated by IF-1, the HPD scans the entries in *AvailableLinkCache,* chooses the appropriate link to connect to (assume it is IF-2 link), and send a MIH_link_Connect request to MIHF to set this connection (L2 soft Handoff). Upon receiving a Link_Up from IF-2, the HPD solicits the NEMO mobility support to perform if required CoA acquisition and registration and tunnel establishment (L3 soft Handoff).

Table 2. Mobile Router Available Links Cache

| MR IF MAC Addr | MAC addr of new PoA | MIH capability | Link Type | Expire Time |
|---|---|---|---|---|
| IF-2 | | | | |
| IF-3 | | | | |

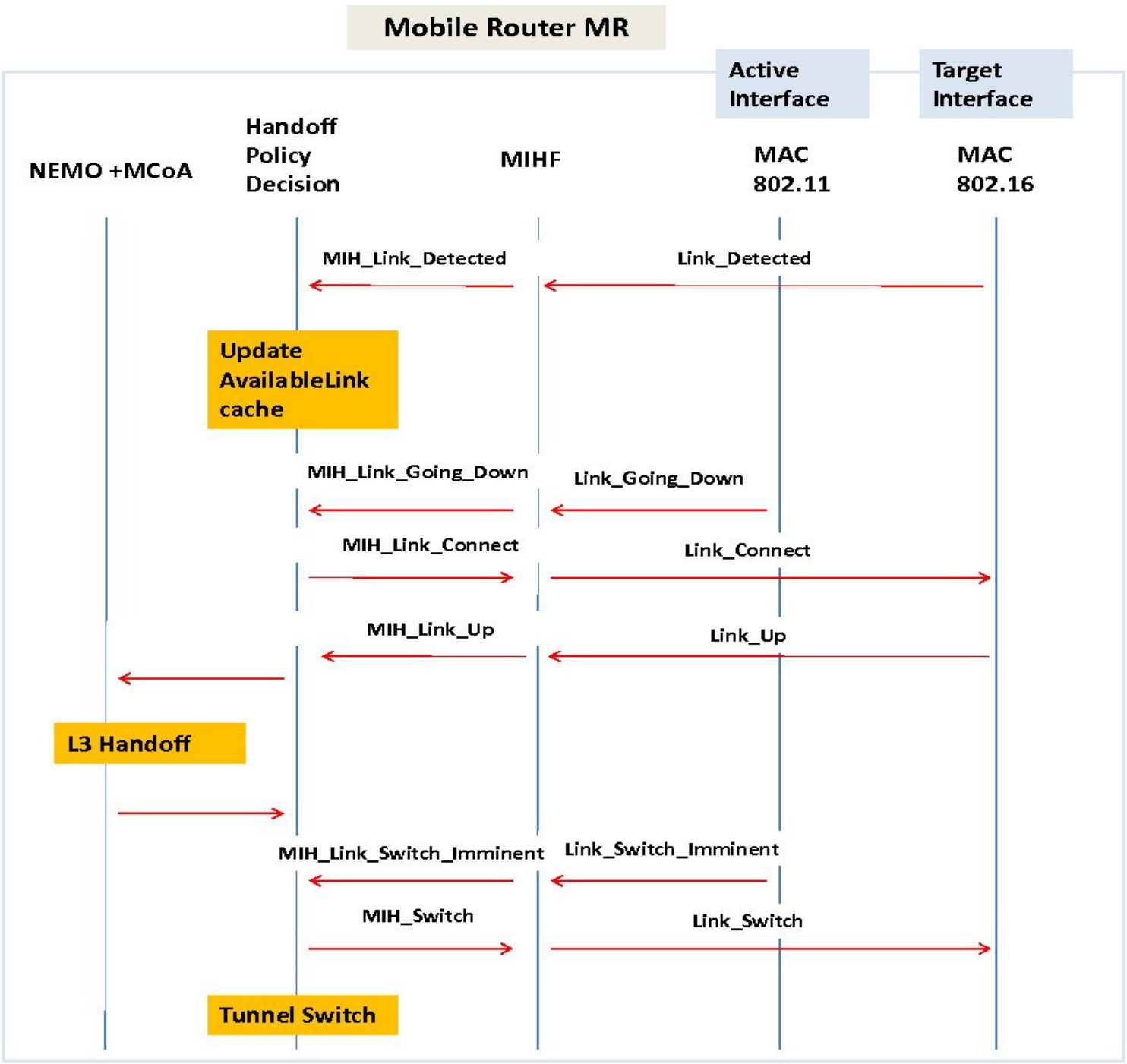

Figure 13. Proposed Handoff preparation and execution procedures

When processing Link_Going_Down event, if the received signal power $\overline{P_{rx}}$ goes up $\alpha_{LGD}.P_{th}$ MIH_Link_Event_Rollback is generated.

To establish a second tunnel between the mobile router (MR) and the home agent (HA), multiple care-of-addresses (MCoA) is used [28]; we modify the binding cache structure of the HA (Table 3) to accommodate multiple binding registrations at the HA. The second established tunnel remains in status "standby" until it is switched to active mode when a Tunnel_Switch_Request message is received from the MR and validated by the HA.

Table 3. Home Agent Binding Cache

| HoA | BID | CoA | Tunnel Status | Expire Time |
|---|---|---|---|---|
| HoA1 | BID1 | CoA1 | active | - |
| HoA1 | BID2 | CoA2 | standby | - |

Note that new available paths (links or tunnels) for the MR are stored at the HPD level also in a cache called *AlternativePathCache* (Table 4 ).
Then if a Link_Switch_Imminent event is generated by IF-1, the HPD scans the *AlternativePathCache* to look for an available alternative path. Depending on "Handoff Type" field in *AlternativePathCache* , the HPD will request only link switching (MIH_Link_Switch) or both link switching and tunnel switching (request to NEMO).

Table 4. Mobile Router Alternative Paths Cache

| Link ID | IF | Handoff Type | CoA | Status | Expire Time |
|---|---|---|---|---|---|
| Link # | IF2 | Horizontal/Vertical | CoA2 | ready | - |

To allow NEMO to perform tunnel switching, we define two new NEMO signaling messages with MH Type = 9 (Tunnel_Switch_Request message, see Figure 14) and MH Type = 10 (Tunnel_Switch_Replay message, see Figure 15) in the Mobility Header of NEMO protocol [2].

| Payload Proto | Header Len | MH Type = 9 | Reserved |
|---|---|---|---|
| Checksum | | Sequence ID | Time |
| HoA | | | |
| BID1 of active tunnel | | BID2 of target tunnel | |
| IPv6 care-of address (CoA) of active tunnel | | | |
| IPv6 care-of address (CoA) of target tunnel | | | |
| options | | | |

Figure 14. Packet Format of Tunnel_Switch_Request message

| Payload Proto | Header Len | MH Type = 10 | Reserved |
|---|---|---|---|
| Checksum | | Sequence ID | Time |
| Replay Code | | | |

Figure 15. Packet Format of Tunnel_Switch_Replay message

After a period time twice the time $T_{LGD}$ from the time a Link_Going_Down event is generated, if neither a Link_Switch_Imminent event nor a Link_Down event is generated, the IF-2 is disconnected, the alternative path is deleted from *AlternativePathCache* and the tunnel is removed from the binding cache at the HA level.
In any case, if a Link_Down event is generated, the HPD takes the decision to switch to an alternative path if available, otherwise to Handoff to an alternative link if available, otherwise to scan for new access networks.

# 6. Simulation Results

The scenario illustrated in Figure 16 was simulated using the NS-2 simulator together with the NIST mobile package to verify and evaluate the extended NEMO model described previously. The network topology is constituted of six nodes using hierarchical addressing, a router (0.0.0), two access routers: the base station 802.11 AR1 (1.0.0) with coverage of 100 m and the base station 802.16 AR1 (2.0.0) with coverage of 1000 m, the mobile router MR (4.1.0) moving at speed 90 Km/h from AR1 cell to AR2 cell, the Home Agent HA (4.0.0) and the correspondent node CN (3.0.0). Link characteristics namely the bandwidth and the delay are shown are also shown on the figure. Simulation time is set to 60 s. A Constant Bit Rate (CBR) traffic stream with a packet size of 768 bytes at 0.016 second intervals is sent from CN to MR. A shadowing model was used for the 802.11 radio link with $\sigma = 4$, $\beta = 3$, a transmit power of 14 dBm and a predefined threshold power $P_{th}$ equal to -75 dBm.

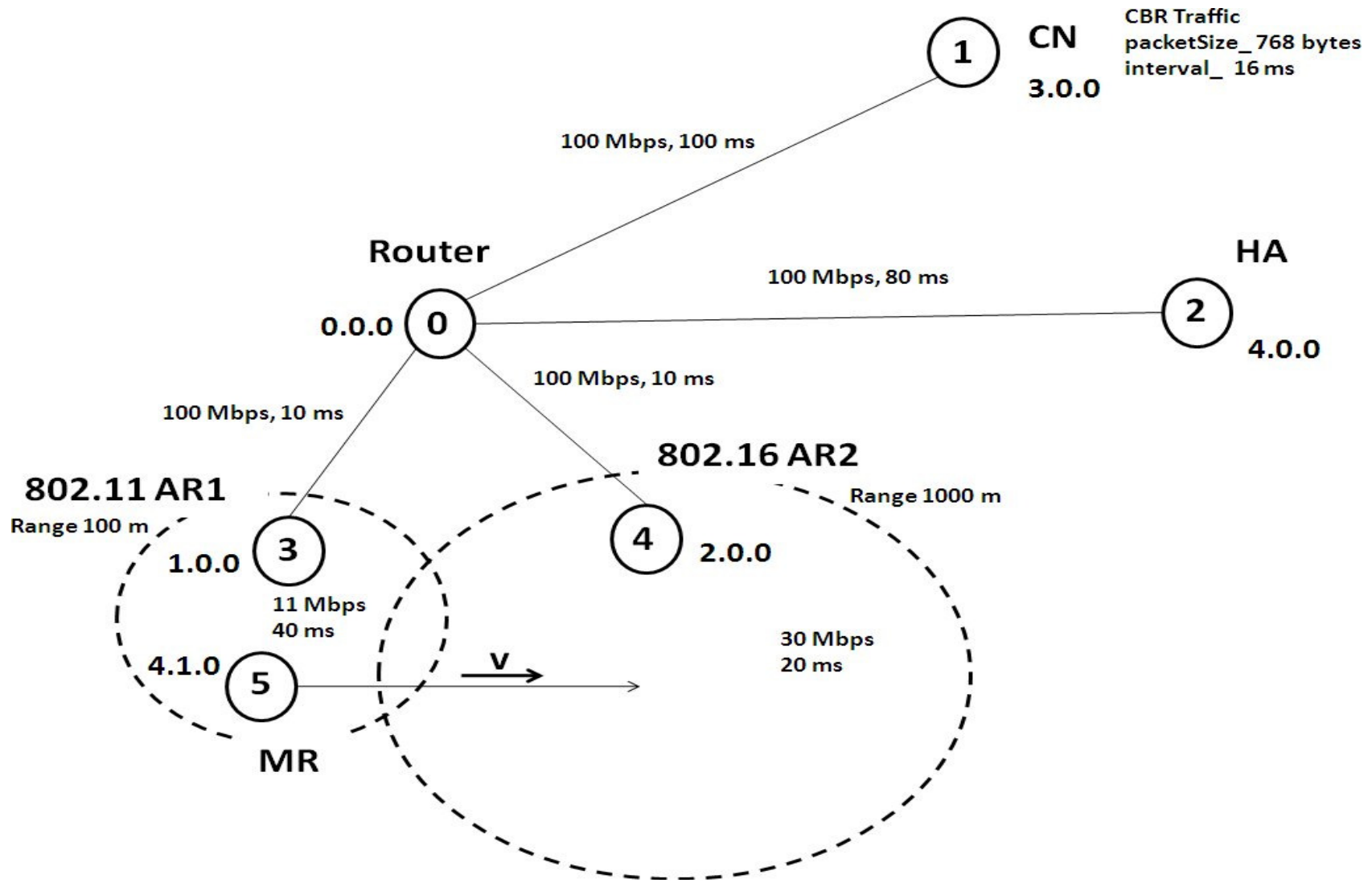

Figure 16. Simulated Network topology

First, we investigate appropriate value for $\delta$ for accurate estimation of received signal power. $\delta$ will largely depend on the amount of signal variation $\sigma$. Figure 17 shows the possible signal strength variations for different $\delta$ values for a shadowing model with $\sigma = 4$. The variation swing can be seen to be quite large without any averaging applied, while a value of $\delta$= 0.1 stabilizes the estimation quite acceptably. It is important to obtain stability to reduce the probability of a ping pong effect. Note that when more averaging is applied ($\delta$= 0.01) the system becomes less responsive to rapid changes.

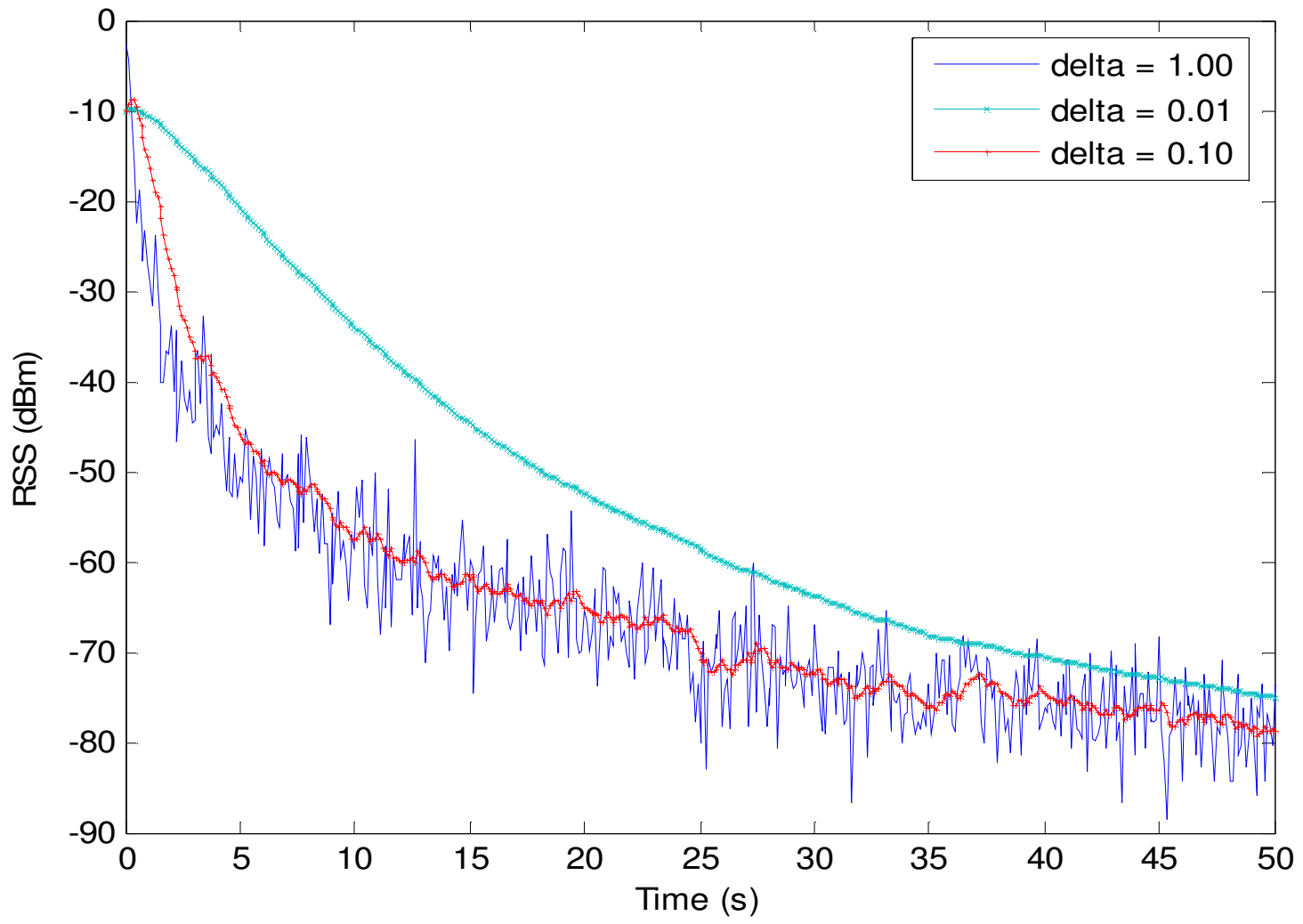

Figure 17. Average received signal strength (RSS) for $\delta$ values of 1, 0.01 and 0.10. $(\sigma = 4, \beta = 3, v = 90\ Km/h)$

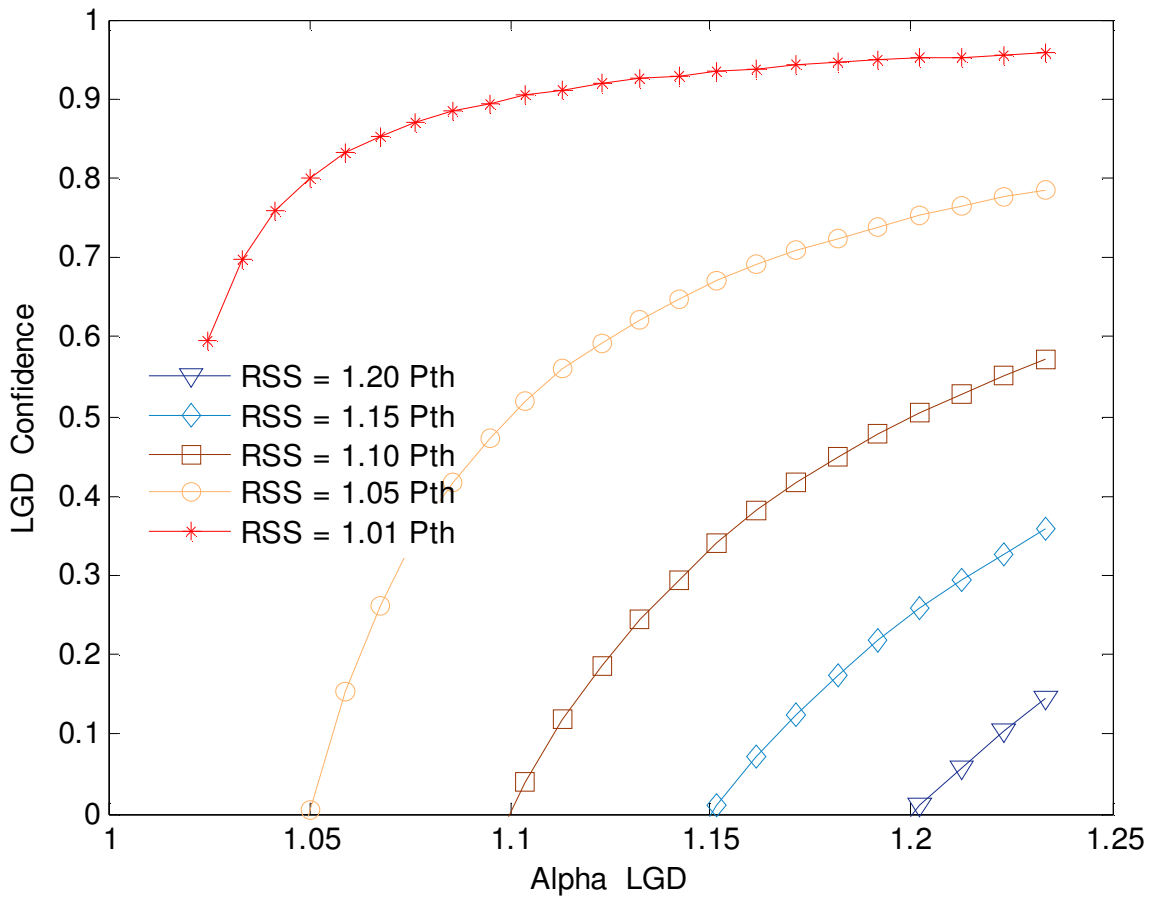

Figure 18. Confidence level for LD event when LGD event is triggered

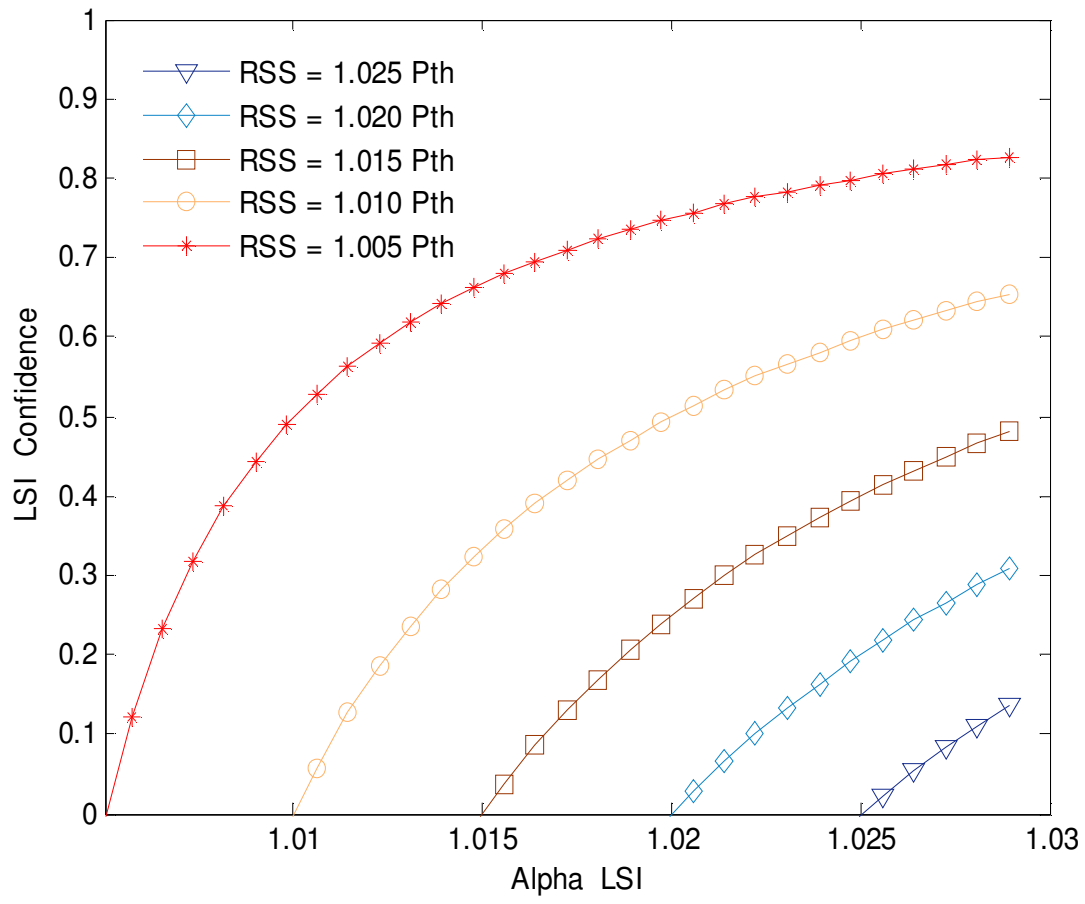

Figure 19. Confidence level for LD event when LSI event is triggered

In Figures 18 and 19 we present the confidence level for link to go down within the specified time interval for respectively LGD and LSI triggers. For a given RSS, the confidence level increases for both LGD and LSI triggers when the corresponding threshold factor increases.

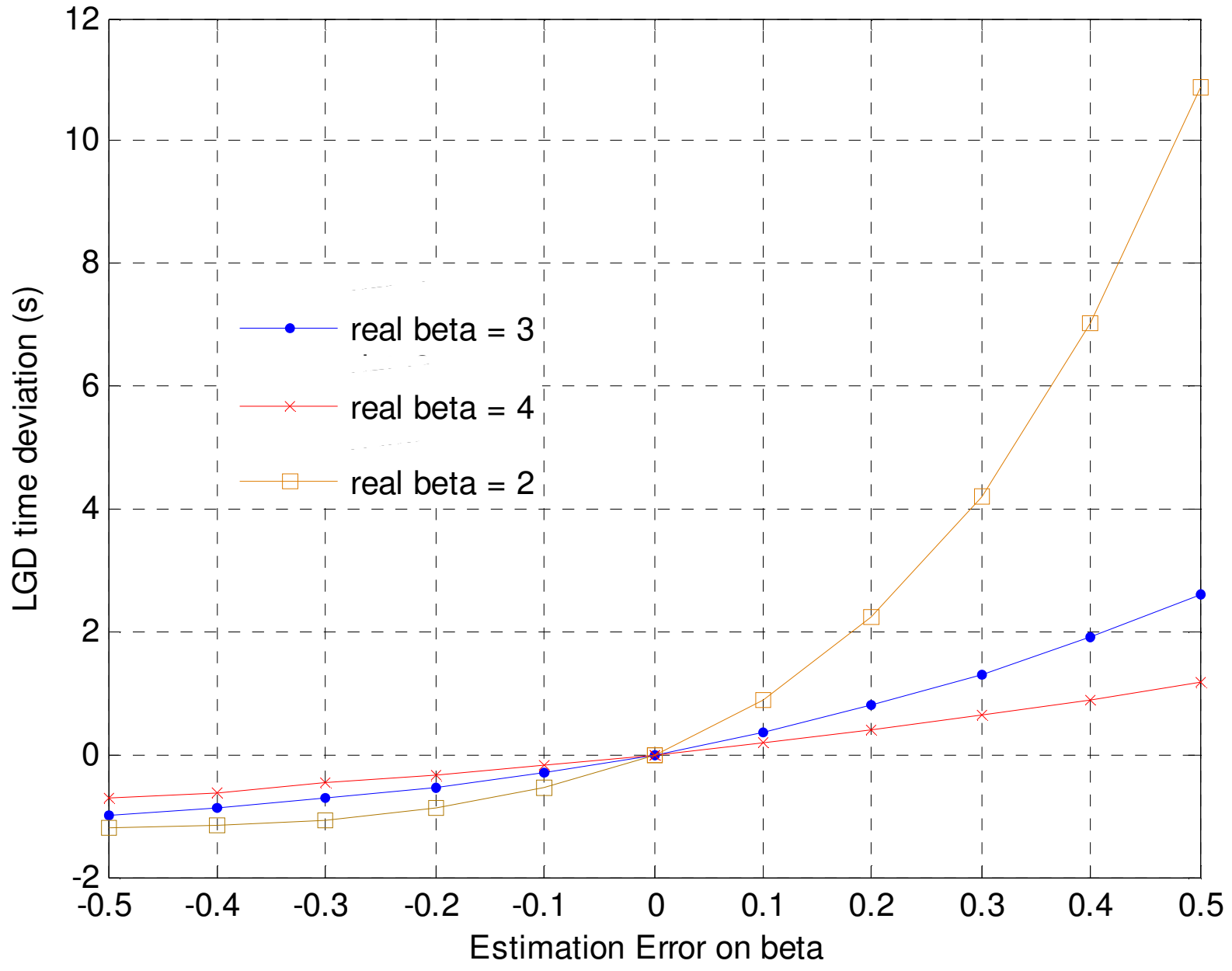

Figure 20. impact of $\beta$ estimation error on $T_{LGD}$

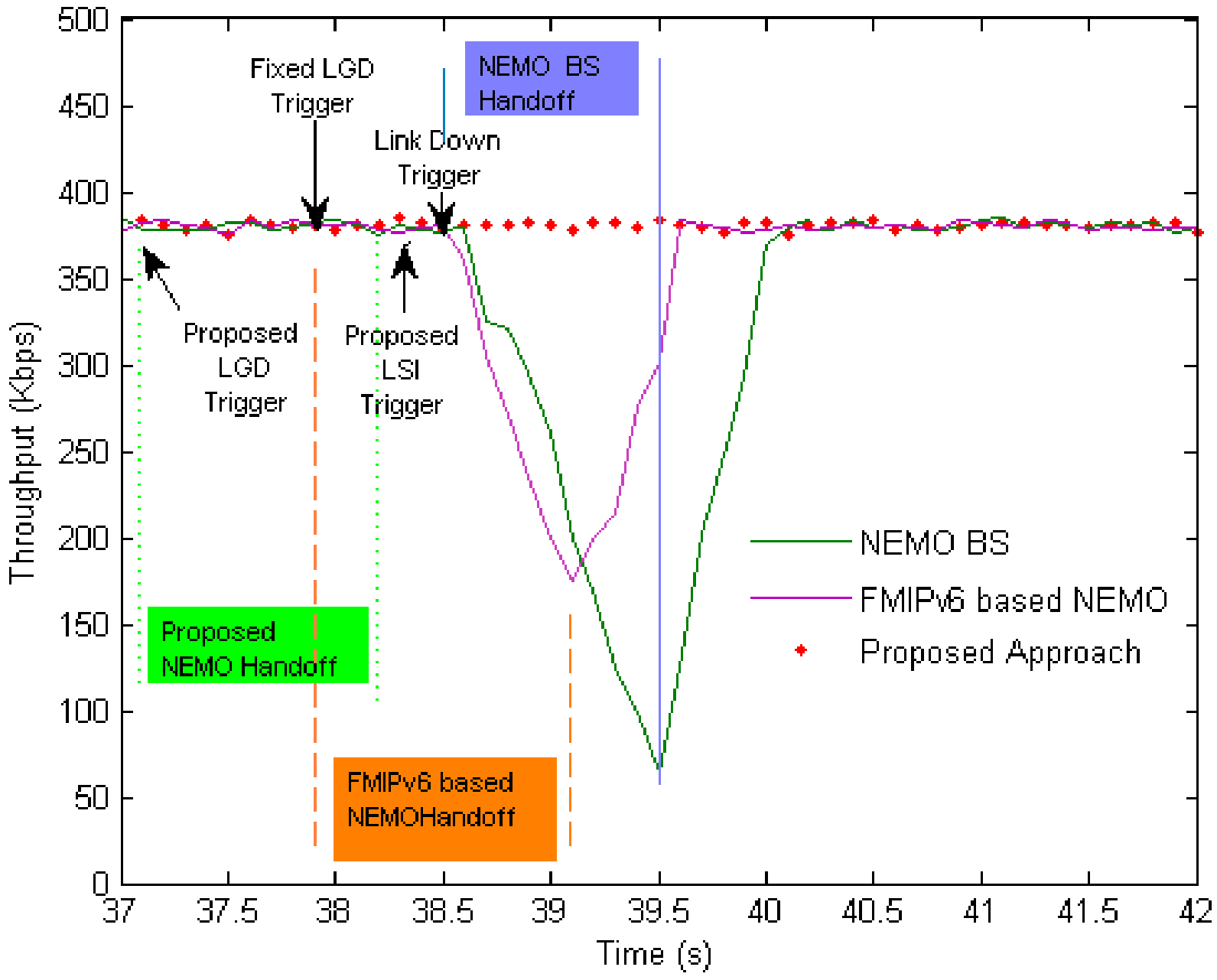

Figure 21. Throughput of received CBR Traffic

We also determined the impact of estimation error on the parameter model $\beta$ on Setting LGD trigger Threshold. The results are shown in Figure 20 when the real path loss model involves a value of $\beta = 2, 3$ or $4$. We notice that positive (negative) error leads to increasing (decreasing) in Handoff anticipation time.

Figure 21 shows the throughput of the CBR traffic at the MR level for the scenario presented in Figure 16. The model was used without $\beta$ estimation error ($\Delta\beta$=0) and a value of $\delta$= 0.1 for RSS estimation. The result is compared with MIPv6-NEMO (Handoff triggered by LD) and FMIPv6-NEMO (Handoff anticipation triggered by LGD with fixed $\alpha_{LGD} = 1.05$). The LD occurs at time 38.512 s. For FMIPv6-NEMO, the LGD event is triggered at 37.893 s. For our proposal, the LGD is triggered at 37.146 s and LSI is triggered at 38.396 s
While MIPv6-NEMO and FMIPv6-NEMO achieve both finite Handoff delay and finite packet loss, our proposal provides seamless connectivity with no Handoff latency and no packet loss.

# 7. CONCLUSIONS

In this paper, we have investigated the combination of multihoming and intelligent soft handoff to achieve seamless connectivity for real time and QoS-sensitive applications in the context of NEMO networks. We addressed the case of (1,1,1) multihomed NEMO model with the assistance of IEEE 802.21MIH services. The proposed Handoff mechanism must be executed before the Link_Down event of the current link. For this purpose, we used LGD trigger (defined by required NEMO Handoff time) for Handoff preparation and LSI trigger (defined by required tunnel switching time) for Handoff anticipation. Our contributions are the design of a new MIH user (HPD: Handoff Policy Decision) for intelligent soft Handoff decisions based on information gathered from surrounding networks, the definition of new MIH service to provide LSI trigger and the extension of the NEMO BS protocol to support tunnel switching when MCoA registration is used. The tests we performed show that our solution makes it possible to achieve a really seamless handover when the suitable model and parameters are chosen. Our proposed Handoff approach is infrastructure independent and can provide both no packet loss and no Handoff delay as well.

# REFERENCES

[1] V. Devarapalli, R. Wakikawa, A. Petrescu, and P. Thubert, (2005), "Network Mobility (NEMO) Basic Support Protocol," Internet Engineering Task Force (IETF), RFC-3963.

[2] D. Johnson, C. Perkins, and J. Arkko, (2004) "Mobility Support in IPv6,"Internet Engineering Task Force (IETF), RFC-3775.

[3] H. Petander, E. Perera, K.C. Lan, A. Seneviratne, (2006). Measuring and improving the performance of network mobility management in ipv6 networks. IEEE Journal on Selected Areas in Communications, 24(9), pp 1671-1681.

[4] V.Vassiliou and Z. Zinonos, (2009) "An Analysis of the Handover Latency Components in Mobile IPv6," Journal of Internet Engineering, vol. 3(1), pp 230-240.

[5] Shayla Islam and al, (2011). Mobility Management Schemes in NEMO to Achieve Seamless Handoff: A Qualitative and Quantitative Analysis. Australian Journal of Basic and Applied Sciences, 5(6) pp 390-402.

[6] Moore N, (2005). Optimistic duplicate address detection for ipv6. IETF draf.

[7] Kempf J, Khalid M, Pentland B, (2004). Ipv6 fast router advertisement. IETF draft.

[8] Q.B Mussabbir, W. Yao, (2007). Optimized FMIPv6 using IEEE 802.21 MIH Services in Vehicular Networks. IEEE Transactions on Vehicular Technology. Special Issue on Vehicular Communications Networks.

[9] C-W Lee, Y-S. Sun and M-C Chen, (2008). HiMIP-NEMO: Combining Cross-layer Network Management and Resource Allocation for Fast QoS-Handovers. Proceedings of the 67th IEEE Vehicular Technology Conference, Singapore.

[10] Z. Yan, H. Zhou and I. You, (2010). N-NEMO: A Comprehensive Network Mobility Solution in Proxy Mobile IPv6 Network. Journal of Wireless Mobile Networks, Ubiquitous Computing, and Dependable Applications, Vol. 1, No. 2/3, pp 52-70.

[11] H. Lin, H. Labiod, (2007). Hybrid handover optimization for multiple mobile routers-based multihomed NEMO networks, in: Proceedings of IEEE International Conference on Pervasive Service, Istanbul.

[12] P. K. Chowdhury, M. Atiquzzaman, and W. Ivancic, (2006). SINEMO: An IP-diversity based approach for network mobility in space. Second International Conference on Space Mission Challenges for Information Technology (NASA SMC-IT), Pasadena, CA, pp 109-115.

[13] Z. Huang, Y. Yang, H. Hu and K. Lin, (2010). A fast handover scheme based on multiple mobile router cooperation for a train-based mobile network Int. J. Modelling, Identification and Control, Vol. 10, No. 3/4, pp 202-212.

[14] G. Jeney, L. Bokor and Z. Mihaly, (2009). GPS aided predictive handover management for multihomed NEMO configurations. 9th International Conference on Intelligent Transport Systems Telecommunications, pp 69 – 73.

[15] A. Mitra, B. Sardar and D. Saha, (2011). Efficient Management of Fast Handoff in Wireless Network Mobility (NEMO). Working paper series WPS No. 671.

[16] S. Pack, X. Shen, J. Mark and J. Pan, (2007). A comparative study of mobility management schemes for mobile hotspots. In Proceedings of IEEE WCNC, pp 3850–3854.

[17] S. Herborn, L. Haslett, R. Boreli, and A. Seneviratne, (2006). Harmony HIP mobile networks. In Proceedings of IEEE VTC 2006-Spring, vol. 2, pp 871–875.

[18] K. Zhu, D. Niyato, P. Wang, E. Hossain, and D. Kim, (2009). Mobility and Handoff Management in Vehicular Networks: A Survey. Wireless Communications and Mobile Computing, Wiley InterScience, pp 1-20.

[19] Y. Y. An, et.al, (2006), “Reduction of Handover Latency Using MIH Services in MIPv6”, in Proc. of the 20th International Conference on Advanced Information Networking and Applications (AINA’06) – Vol.02, pp 229-234.

[20] R. Koodli, et al., (2008), ”Mobile IPv6 Fast Handovers”, RFC 5268, Internet Engineering Task Force.

[21] M. Woo, H. Lee, Y. Han and S.Min, (2010). A Tunnel Compress Scheme for Multi-Tunneling in PMIPv6-based Nested NEMO. International Journal of Wireless & Mobile Networks (IJWMN) Vol.2, No.4, pp 60-69.

[22] Details for North Americ Internet Traffic Report. http://www.internettrafficreport.com/namerica.htm

[23] IEEE 802.21-2008, Media Independent Handover Services.

[24] S. Woon, N. Golmie, A. Sekercioglu (2006). Effective Link Triggers to Improve Handover Performance. Proceedings of 17th Annual IEEE Symposium on Personal, Indoor, and Mobile Radio Communications (PIMRC'06), Helsinki, Finland, pp 11-14.

[25] S. J. Yoo, D. Cypher, and N. Golmie, (2007), "LMS predictive link triggering for seamless handovers in heterogeneous wireless networks," in Proc. MILCOM, Orlando, FL, Oct. 28–30, pp 1–7.

[26] Theodore S. Rappaport (2002). Wireless Communication: Principles and Practice. Personal Education International.

[27] C. Ng, E. Paik, T. Ernst, and M. Bagnulo, (2007), "Analysis of Multihoming in Network Mobility Support," IETF, RFC 4980.

[28] R. Wakikawa, V. Devarapalli, G. Tsirsis and T. Ernst and K. Nagami, (2009), "Multiple Care-of Addresses Registration," IETF, RFC 5648.